\documentclass[manuscript]{geophysics}

\begin{document}

\title{Progressive transfer learning for low frequency data prediction in full waveform inversion}

\renewcommand{\thefootnote}{\fnsymbol{footnote}}

\ms{manuscript v1.1} 

\address{
\footnotemark[1]Advanced Geophysical Technology, Inc. \\ 14100 Southwest Freeway, Suite 110, Sugar Land, Texas 77478 \\ Email:wenyi.hu@agtgeo.com \\
\footnotemark[2]University of Houston, Houston, Texas 77204
}

\footer{Deep learning for low frequency prediction}
\lefthead{Hu, Jin, Wu \& Chen}
\righthead{Low frequency prediction by deep learning}

\author{Wenyi Hu\footnotemark[1], Yuchen Jin\footnotemark[2], Xuqing Wu\footnotemark[2],and Jiefu Chen\footnotemark[2]}

\maketitle

\begin{abstract}
For the purpose of effective suppression of the cycle-skipping phenomenon in full waveform inversion (FWI), we developed a Deep Neural Network (DNN) based approach to predict the absent low-frequency components by exploiting the hidden relation connecting the low-frequency data and the high-frequency data implicitly through the subsurface geological and geophysical properties. In order to efficiently solve this challenging nonlinear regression problem, two novel strategies were proposed to design the DNN architecture and to optimize the learning process: 1) Dual Data Feed structure; 2) Progressive Transfer Learning. With the Dual Data Feed structure, not only the high-frequency data, but also the corresponding Beat Tone data are fed into the DNN to relieve the burden of feature extraction, substantially reducing the network complexity and the training cost. The second strategy, Progressive Transfer Learning, enables us to unbiasedly train the DNN using a single training dataset that is generated by an arbitrarily selected velocity model. Unlike other established deep learning approaches where the training datasets are fixed, within the framework of the Progressive Transfer Learning, the training velocity model and the associated training dataset continuously evolve in an iterative manner by gradually absorbing more and more reliable subsurface information retrieved by the physics-based inversion module, progressively enhancing the prediction accuracy of the DNN and propelling the velocity model inversion process out of the local minima. The Progressive Transfer Learning, alternatingly updating the training velocity model and the DNN parameters in a complementary fashion toward convergence, saves us from being overwhelmed by the otherwise tremendous amount of training data, and avoids the underfitting and biased sampling issues. The numerical experiments validated that, without any \emph{a priori} geological information, the low-frequency data predicted by the Progressive Transfer Learning are sufficiently accurate for an FWI engine to produce reliable subsurface velocity models free of cycle-skipping-induced artifacts.
\end{abstract}

\section{Introduction}

Full waveform inversion (FWI), mathematically built on nonlinear optimization algorithms, is an advanced seismic processing technology for high resolution subsurface velocity model building through a data-fitting procedure. Due to the extremely large number of unknowns in industrial-sized velocity model building projects, most FWI methods employ the gradient-based local optimization algorithms that often become stagnant at a local minimum if the inversion process is initiated at relatively high frequencies or the starting velocity model is not sufficiently close to the true model.  Because reliable low frequency (LF) components below 5 $Hz$ do not practically exist in most acquired seismic datasets and accurate starting velocity models are usually not available, FWI often suffers from the local minimum issue, or equivalently the cycle-skipping phenomenon, observed as strong artifacts and wrong velocity update contaminating the reconstructed subsurface velocity models. Sometimes, these cycle-skipping-induced artifacts are hard to identify from the real seismic reflection events and may eventually lead to improper interpretation of the subsurface geological structures, preventing FWI from being widely applied in subsurface hydrocarbon exploration.

In recent years, there was a surge of research efforts devoted to overcome this challenge, i.e., eliminating or suppressing the cycle-skipping phenomenon without physically acquiring LF data. Many of these research works can be classified into the following categories \cite[]{Hu_etal2018}: 1) scattering angle based filtering methods \cite[]{Alkhalifah2014}. This category of methods was developed based on the observation that the subsurface reconstruction resolution is highly dependent on the subsurface scattering angle. 2) FWI with extended velocity model space \cite[]{Symes2008,Fu_and_Symes2015,Fu_and_Symes2017}. These approaches introduce an additional dimension into the unknown velocity model and gradually enforce the shift in this additional dimension towards zero. 3) FWI with time shift minimization \cite[]{Xu_etal2012, Ma_and_Hale2013}, where a traveltime difference minimization is implemented to correct the kinematic information for the cycle-skipping prevention. 4) FWI with synthesized low frequency data \cite[]{Shin_and_Cha2008, Wu_etal2014, Li_and_Demanet2016}. These methods apply various nonlinear operators to map the high frequency (HF) data to the low frequency (LF) domain. 5) FWI resolving phase ambiguity \cite[]{Hu2014, Choi_and_Alkhalifah2015}. All of these research efforts helped mitigating the cycle-skipping issue to some extent. However, till today, this challenging problem has not been completely solved.

More recently, embracing the power of machine learning and data analytics, some researchers resort to the Deep Neural Network (DNN) approaches to predict the absent LF data by learning the underlying nonlinear relationship between the HF components and the LF components \cite[]{Jin_etal2018, Sun_and_Demanet2018, Ovcharenko_etal2018}. While the early stage results of these pure data driven methods are encouraging, some major technical issues related to accuracy and efficiency need to be addressed before we hit a bottleneck in production. Some of the main challenges are 1) How to design the training datasets to ensure an unbiased training process? 2) How to reduce the amount of training data to a more computationally manageable level without sacrificing the LF prediction accuracy? 3) How to design quality control measures to monitor the training process and to quantitatively evaluate the reliability of the prediction results?

In this work, after conducting a feasibility study of the LF components synthesis from the HF data, we proposed and developed a novel DNN based approach to address these issues. There are two unique features differentiating our DNN approach from other existing methods: 1) the Dual Data Feed network structure featuring two input branches receiving the HF data and the corresponding Beat Tone data simultaneously, thus partially relieving the network from the heavy burden of feature extraction; 2) the Progressive Transfer Learning strategy, a striking feature integrating the deep learning (DL) module and the physics-based module seamlessly to iteratively enhance the LF data prediction accuracy. With this unique Progressive Transfer Learning strategy, only a single set of training data is required to initiate and to complete the network training process. Unlike most conventional DNN approaches where the training datasets are always fixed during the network training process, we update the training velocity model after each learning cycle to inject a progressively improving training dataset to the DL module. Consequently, the network training cost is substantially reduced and the LF data prediction accuracy is significantly improved, while the quality of the training process is quantitatively monitored. This unique feature implies its great potential in large industrial-sized projects while other machine learning methods tend to fail due to the limited availability of training data, the computational efficiency bottleneck, and the algorithm convergence issue.

\section*{Role of Low Frequency Components in FWI}
Without losing generality, the cost function of a standard FWI can be posed in the frequency domain as
\begin{equation}
\label{eqn1}
C(\mathbf{v})=\sum_{f=1}^{N_f} \sum_{s=1}^{N_s} \sum_{r=1}^{N_r} \|\mathbf{S}_{r,s,f}(\mathbf{v})-\mathbf{M}_{r,s,f}\|^{2},
\end{equation}
where $S$ are the simulation data, $M$ are the measurement data, and $\mathbf{v}$ are the subsurface seismic velocities to be reconstructed. The subscripts $f$, $s$, and $r$ represent the indices of the frequency, source, and receiver.$N_s$, $N_r$, and $N_f$ are the number of sources, receivers, and frequencies, respectively. The gradient of the cost function with respect to the seismic velocity can be computed by performing a residual back-propagation procedure \cite[]{Pratt_etal1998}, which is used to update the subsurface velocity model and minimize the discrepancy between the simulation data and the measurement data. The velocity model update is carried out iteratively until the data misfit is within a predefined error tolerance \cite[]{Hu_etal2009}. Unfortunately, as a nonlinear optimization problem, minimization of the cost function (\ref{eqn1}) is not always successful. The gradient-based algorithm often converges to a local minimum and the data misfit may not be continuously reduced to the error tolerance, depending on the nonlinearity level of the problem. The important role played by low frequency data can be perceived in Figure~\ref{fig:figure_1}, a diagram showing a simple one-variable nonlinear optimization problem. As observed in Figure~\ref{fig:figure_1}, at a high frequency (HF), a starting velocity model is often not in the basin of attraction of the global minimum; hence the algorithm tends to converge to a local minimum $V_L$ instead of the global minimum $V_T$. An effective strategy for tackling this issue is to initiate the FWI inversion at a lower frequency (if the lower frequency component is available in the acquired seismic data), where the non-convexity nature of the cost function is alleviated and the global minimum $V_A$ is obtained at the low frequency regime. After that, the FWI algorithm switches back to the high frequency regime with a better starting velocity model $V_A$ residing in the basin of attraction, securing the convergence of the gradient-based algorithm at the true solution $V_T$. Based on these analyses, the notorious local minimum issue, commonly referred to as the cycle-skipping phenomenon in FWI, can be successfully prevented if the optimization is performed at a sufficiently low frequency because the nonlinearity of FWI reduces as the frequency decreases. From this perspective, our task of cycle-skipping suppression can be converted to solving an equivalent problem - the LF prediction from the recorded LF seismic data, which is a challenging nonlinear regression problem.

\renewcommand{\figdir}{Fig} 
\plot{figure_1}{width=0.8\textwidth}
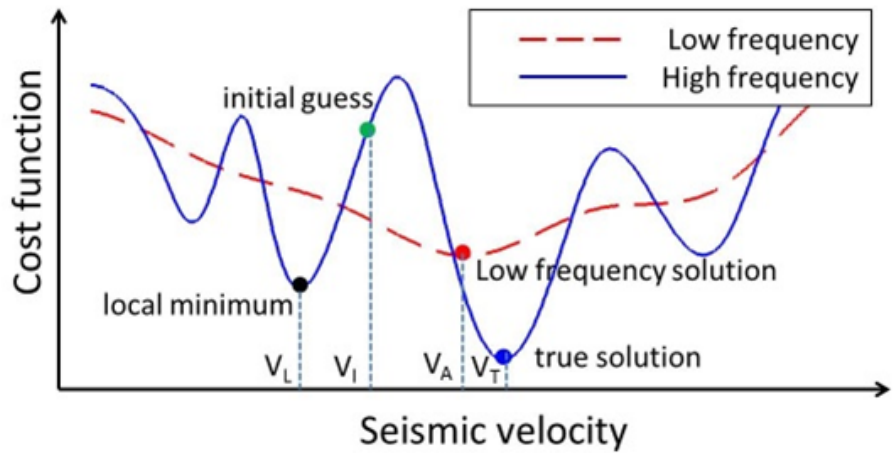
{A diagram showing a one-variable nonlinear optimization problem. $V_L$: local minimum; $V_I$: starting velocity model; $V_A$: approximate solution obtained at low frequency; $V_T$: true solution.}

\section*{Feasibility Study of Low Frequency Prediction}

Before delving into a DNN to perform this challenging task, the feasibility of the LF data prediction needs to be investigated. Is it possible to recover the absent LF seismic data from the acquired HF seismic data? At first glance, this question appears to be a pure signal processing topic but, as a matter of fact, the answer not only lies in the data, but also relies on the specific subsurface geological and geophysical environment where the data are acquired.

\multiplot{1}{figure_2a,figure_2b}{width=0.8\textwidth}
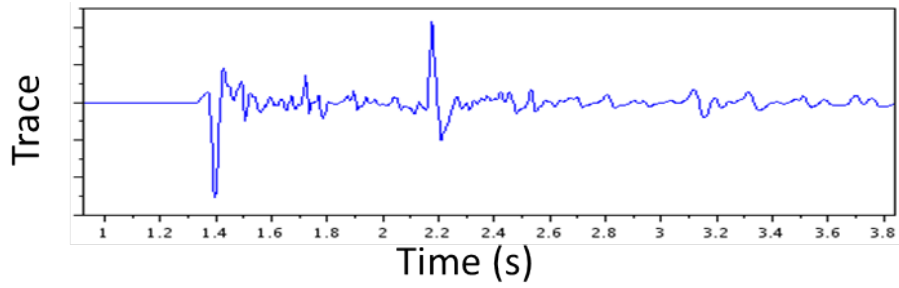
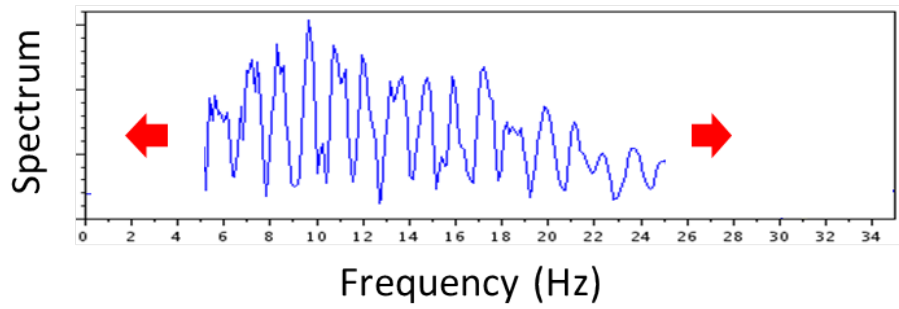
{a) a seismic trace with sparse reflection events; b) spectrum of the trace with frequency components below 5 $Hz$ and above 25 $Hz$ abandoned.}

For a general random signal, prediction of the LF information from the HF components is essentially impossible because there is no meaningful relationship between the different frequency bands. However, an important fact often neglected is: seismic data are not random signals. Instead, seismic data are the earth responses to the wideband excitation sources, recorded as the coherent bandlimited signals. This fact suggests an underlying nonlinear tunnel connecting the LF components and the HF components of the seismic data. This nonlinear relationship, whose explicit mathematical form is often unavailable, is unambiguously determined by the geological settings and the geophysical properties of the subsurface media. Here, as a feasibility study, we aim to validate this statement empirically by conducting a seismic bandwidth extension experiment using the sparsity constrained inversion \cite[]{Liang_etal2017}. The input seismic trace for this experiment is plotted in Figure~\ref{fig:figure_2a} and the corresponding frequency spectrum is shown in Figure~\ref{fig:figure_2b}, from which the components below 5 $Hz$ are intentionally removed. On the other hand, the information above 25 $Hz$ is also abandoned due to its relatively low signal-to-noise ratio (SNR). The energy spanning the frequency range between 5 $Hz$ and 25 $Hz$ is then extracted to reconstruct the LF data below 5 $Hz$. For this specific example, based on the reflection characteristics observed in Figure~\ref{fig:figure_2a}, a reasonable assumption can be made that the subsurface reflection events are predominantly sparse, and thus the earth can be depicted by a limited number of discrete reflectors. Consequently, the received seismogram after the deconvolution operation consists of a series of impulses, whose spectrum is mainly contributed by the summation of the frequency domain harmonics corresponding to these time domain impulses. This bandwidth extension task (or equivalently, the LF and HF component prediction) is cast as an inverse problem in the frequency domain with the sparsity constraints defined in the time domain:
\begin{equation}
\label{eqn2}
argmin\|\mathbf{\tilde{d}}-\mathbf{Ac}\|^2_2+\lambda \|\mathbf{c}\|_1,
\end{equation}
where $\mathbf{\tilde{d}}$ is the truncated length-M frequency spectrum (5 $Hz$ $\leq f \leq$ 25 $Hz$) of the input seismic trace, $\mathbf{c}$ is the length-$N$ coefficient sequence, with $M < N$, and the matrix $\mathbf{A}$ is defined as
\begin{equation}
\label{eqn3}
A_{m,n} = exp(j\frac{2\pi}{N}mn), 0\leq n \leq N-1.
\end{equation}
Only when $N = M$, the coefficients $\mathbf{c}$ satisfying $\mathbf{\tilde{d}}=\mathbf{Ac}$ are uniquely determined. However, in scenarios where $N > M$, the non-uniqueness nature of the inverse problem can be overcome by the introduction of the $L_1$ regularization in (\ref{eqn2}). The sparsity-constrained minimization problem (2) is solved by the split variable augmented Lagrangian shrinkage algorithm (SALSA) \cite[]{Afonso_etal2010, Afonso_etal2011} and an approximate solution of $\mathbf{c}$ can be found to reconstruct the LF components below 5 $Hz$. The phase of the predicted LF data is in good agreement with the input data as observed in Figure~\ref{fig:figure_3}, strongly suggesting the feasibility of the LF prediction provided that the subsurface structural sparsity assumption can be justified. We are obliged to mention that, in this numerical experiment, the amplitudes of the predicted LF signals deviate from the input data, partially due to the unknown density, attenuation properties, source wavelet estimation error, imperfect impulse responses of the earth, noise contamination, and many other unknown and uncertain factors. For this reason, the methodology proposed in this work is only intended for the LF phase information retrieval.

\plot{figure_3}{width=0.8\textwidth}
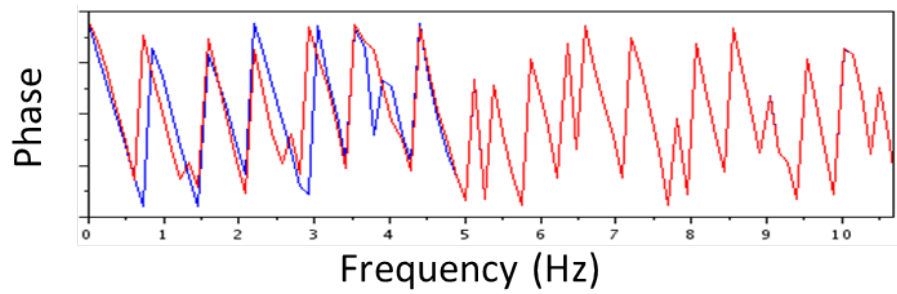
{Comparison between phase spectra of input data (blue) and predicted LF data (red).}

\section*{Deep Learning Approach to Low Frequency Prediction}

Inspired by the conventional bandwidth extension technology that has been widely applied to geological interpretation and reservoir characterization, we aim to extend this trace-by-trace approach to more general scenarios to be suited for FWI. The subsurface reflector sparsity is a critical element in the conventional bandwidth extension algorithm because it is the frequency domain harmonic feature emphasized by the structural sparsity makes the bandwidth extension possible and affordable \cite[]{Zhang_and_Castagna2011}. Although the subsurface structural sparsity assumption is defensible for most qualitative geological interpretation applications, it can be hardly maintained for FWI applications unless the geological environment under investigation is oversimplified. As a direct result of non-sparse reflector structure, the single-trace approach employed in the previous example often gives an unreliable or meaningless solution because the optimization problem (2) becomes an underdetermined system. For this reason, the adaption of the conventional bandwidth extension approach to the requirements of FWI is not trivial.

Even if the sparsity condition of the subsurface structure does not hold for most FWI projects, the strong connection between the LF and HF components still exists but the difficulty level of LF prediction is dramatically increased. Theoretically, a natural and valid strategy for generalizing the conventional bandwidth extension is to fully exploit this connection by bringing in the adjacent seismic traces to participate in the LF data reconstruction process as the extra implicit constraints to the optimization problem.  This strategy often leads to an extremely large-sized nonlinear inverse problem beyond the available computation capability. Alternatively, \emph{a priori} subsurface geological information can be incorporated into the inversion process to alleviate the non-uniqueness of the inverse problem, but this strategy is not straightforward and often unmanageable with the concern over the availability and reliability of the subsurface information. In this context, we resort to the pure data-driven approach - machine learning.

Machine learning (ML) is an algorithm or a mathematical model that is used to analyze data, recognize the numerous patterns in the data that is impossible for a human being to identify exhaustively, learn the underlying nonlinear relationship between the input and the output automatically, and then predict the value of new output data given the new input data points. Deep learning (DL) or DNN is a subset of machine learning methods established on the architecture featuring multiple layers of artificial neural networks to make predictions and decisions by automatically extracting the unique features buried in the data. Deep learning is especially suitable for this work because it is infeasible to manually extract the numerous relevant sets of features to accurately capture the underlying nonlinear relationship between the LF and HF components of the seismic data.

\plot{figure_4}{width=0.8\textwidth}
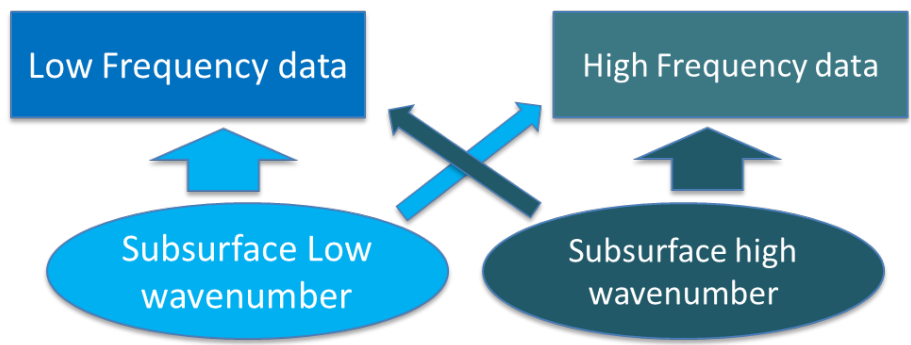
{Diagram of relationship between seismic data and wavenumber components of subsurface structures.}

Theoretically, a properly designed DNN for LF prediction is able to automatically and successfully perform the feature extraction task as long as the connection between the LF data and the HF data is solid and stable. Figure~\ref{fig:figure_4} is a diagram qualitatively sketching this connection, where we note that the HF data are dominantly contributed from the subsurface high wavenumber structures while only being vaguely connected to the low wavenumber structures via far offset source/receiver geometry setting. This fact directly affects the feature extraction capability and the effectiveness of the DNN for LF prediction. In order to relieve the burden of feature extraction on the network, we need to somehow amplify the connection between the low wavenumber structure and the HF data to build an unimpeded path linking the LF data, the low wavenumber components of the subsurface structures, and the HF data.

%
\multiplot{1}{figure_5a,figure_5b,figure_5c}{width=0.8\textwidth}
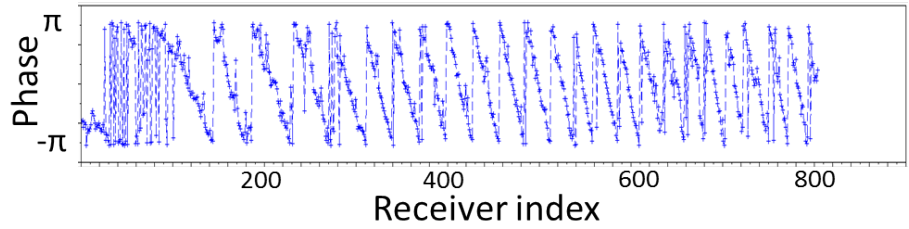
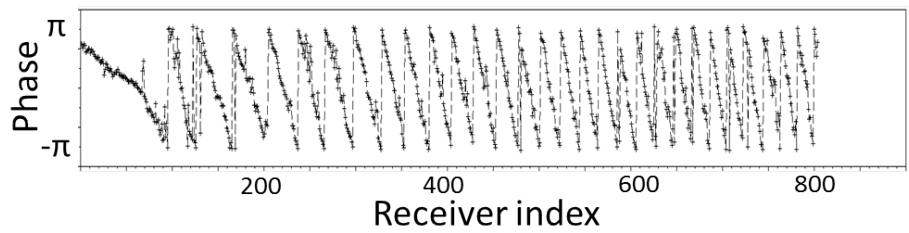
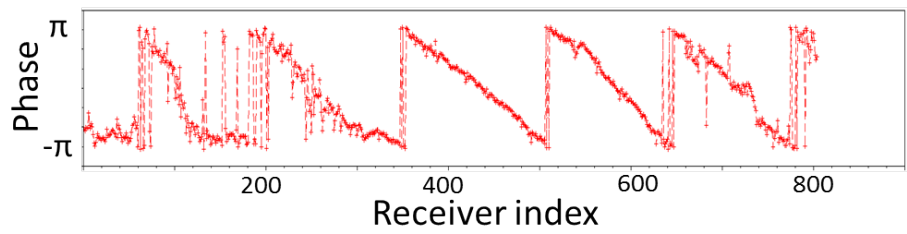
{Comparison between high frequency data and the corresponding Beat Tone data. a) 6 $Hz$; b) 7 $Hz$; c) Beat Tone data derived from 6 $Hz$ and 7 $Hz$ data.}
\plot{figure_6}{width=0.8\textwidth}
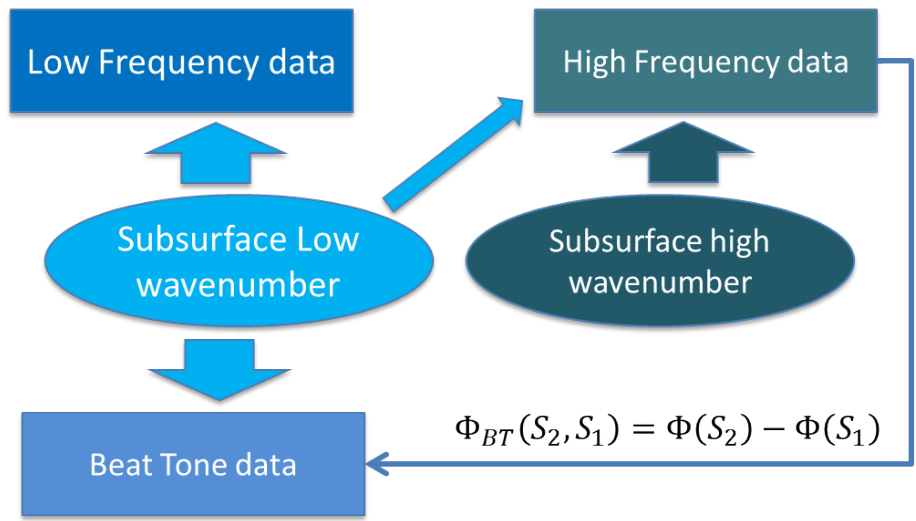
{Diagram of relationship between high frequency data, low frequency data, Beat Tone data, and wavenumber components of subsurface structures.}
To reach that goal, we introduce the Beat Tone data as the second input \cite[]{Hu2014} into the DNN. The Beat Tone technique was developed to suppress the cycle-skipping phenomenon in FWI by amplifying the low wavenumber information buried in the HF data, under the inspiration of the interference Beat Tone, an acoustic phenomenon commonly used by musicians for tuning check. The Beat Tone method utilizes two seismic datasets extracted at two slightly different high frequencies to implicitly reduce the number of phase wrapping occurrences, generating a dataset showing a slow spatial phase variation pattern similar to a true low frequency dataset. A shot gather of 6 $Hz$ data (Figure~\ref{fig:figure_5a}) and 7 $Hz$ data (Figure~\ref{fig:figure_5b}) are compared with the corresponding Beat Tone data in Figure~\ref{fig:figure_5c} to demonstrate that the Beat Tone data is an approximation to 1 $Hz$ data in terms of phase wrapping behavior, although some remaining high wavenumber energy is observed in Figure~\ref{fig:figure_5c}. The calculation of Beat Tone data is straightforward
\begin{equation}
\label{eqn4}
\mathbf{\Phi_{BT}(S_2,S_1)=\mathbf{\Phi(S_2)-\Phi(S_1)}}
\end{equation}
where $\mathbf{\Phi_{BT}}$ is the Beat Tone phase and $\mathbf{S_1}$ and $\mathbf{S_2}$ represent two frequency domain datasets extracted at the frequencies $f_1$ an $f_2$ with $\Delta f=f_2-f_1 \ll f_1$  and $f_2$. With the introduction of Beat Tone data into the DNN as the second input, we establish a solid route connecting the LF components and the HF components through the subsurface low wavenumber structures, as sketched in Figure~\ref{fig:figure_6}. Intuitively, the similarity between the Beat Tone data and the true LF data may help us reduce the complexity of the network architecture without sacrificing the data prediction accuracy of the network. Based on this hypothesis, we designed a DNN that features a Dual Data Feed structure to receive both HF data and Beat Tone data simultaneously via two separate branches A and B sharing the same structure as depicted in Figure~\ref{fig:figure_7}. The outputs of the two branches merge at a stem network to mix the learned features and eventually be mapped to the LF phase prediction. Basically, any convolutional neural network (CNN) could be employed to fill in the frame shown in Figure~\ref{fig:figure_7}. After testing several candidate networks, the inception convolutional network \cite[]{Szegedy_etal2017} was selected and modified to be suited for this nonlinear regression problem because of its high performance in both efficiency and accuracy. Unlike most traditional DNNs stacking deeper and deeper convolution layers to tackle more and more challenging tasks while overfitting and enormous computational cost become the bottlenecks, the inception network on the other hand, uses a variety of novel strategies, such as convolution factorization methods, to enhance the overall performance. Figure~\ref{fig:figure_8a} shows the structure of our 1D inception block adapted from the inceptV4 network, which consists of multiple parallel branches with different convolution depths to capture both local and global patterns in the input data. In Figure~\ref{fig:figure_8a}, the convolution block with the total depth of $D$ is composed of $D+2$ parallel branches, while the depth $d$ associated with the individual branch varies from $-1$ to $D$.  When $d = -1$, there are only a pooling layer and a $1\times1$ convolution layer in the branch. If $d = 0$, the branch only has a projection layer to output a combination of the original features in the data. If $d > 0$, then there are $d$ convolution layers after the projection, resulting in an output with the length of $2x(d-1)x(N/2)+N$. The other basic unit illustrated in Figure~\ref{fig:figure_8b} is the deconvolution block, which up-samples the input layers by the size of the stride. The structure of the regressor is shown in Figure~\ref{fig:figure_9}, where all the individual inception blocks have the same depth of $2$ and the depth of the entire network is $41$. A $tanh$ function is adopted as the nonlinear activation function, which maps the output to values in between -1 and 1. The training process of the network can be set up as an optimization problem minimizing the following loss function
\begin{equation}
\label{eqn5}
\mathbf{min}_\Theta\|f(\mathbf{S_{HF}},\mathbf{S_{BT},\mathbf{\Theta}})-\mathbf{S_{LF}})\|_2^2,
\end{equation}
where the network output is represented by the function $f(\mathbf{S_{HF}},\mathbf{S_{BT}}, \mathbf{\Theta)}$. $\mathbf{S_{HF}}$ and $\mathbf{S_{BT}}$ are the training datasets, HF data and the corresponding Beat Tone data, injected into the two receiving branches of the network. Given the ground truth $\mathbf{S_{LF}}$ (LF training data), the parameters of the network $\mathbf{\Theta}$ can be solved.
\plot{figure_7}{width=0.8\textwidth}
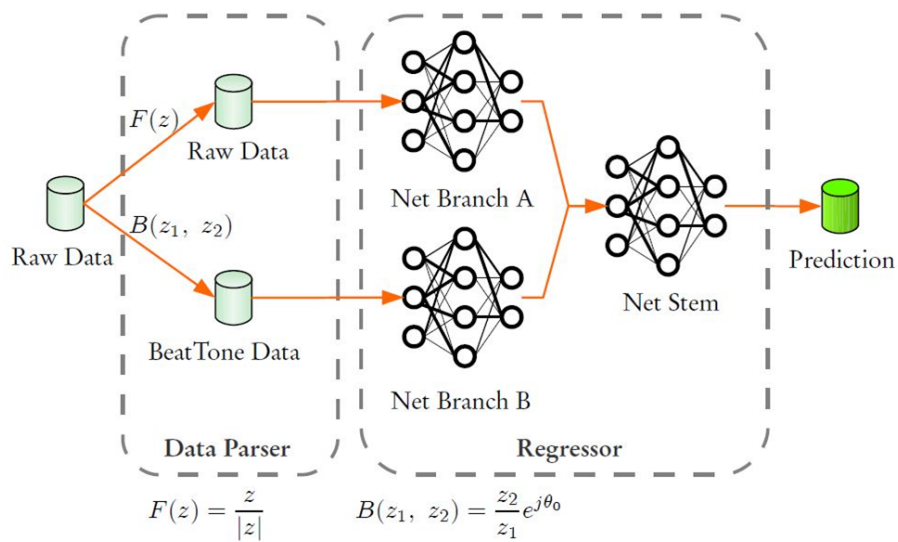
{Dual Data Feed structured network.}
\multiplot{1}{figure_8a,figure_8b}{width=0.8\textwidth}
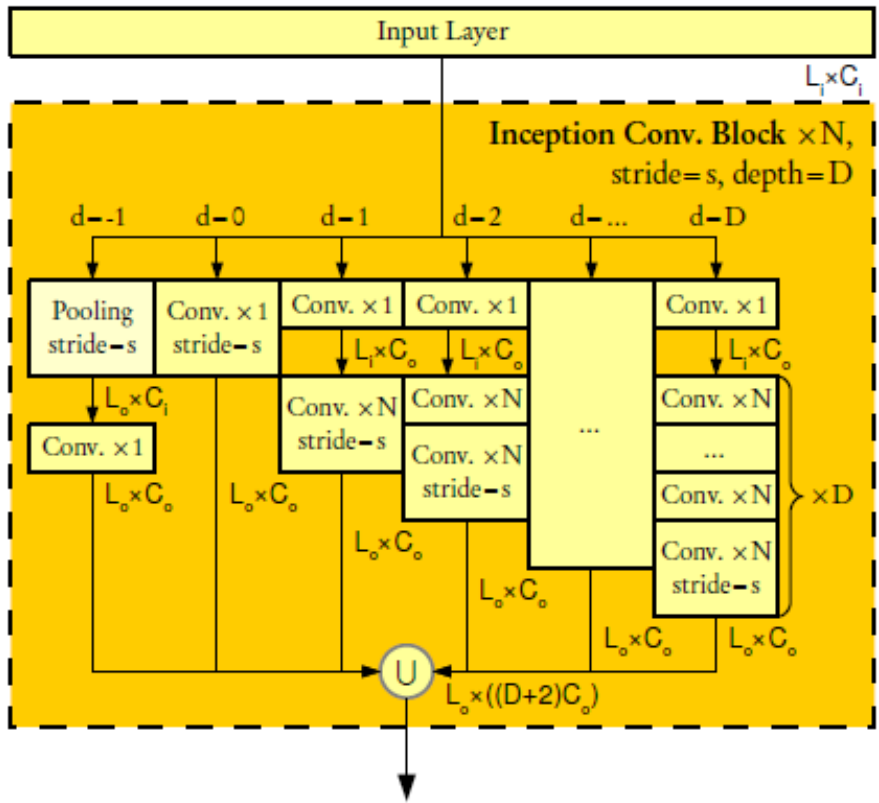
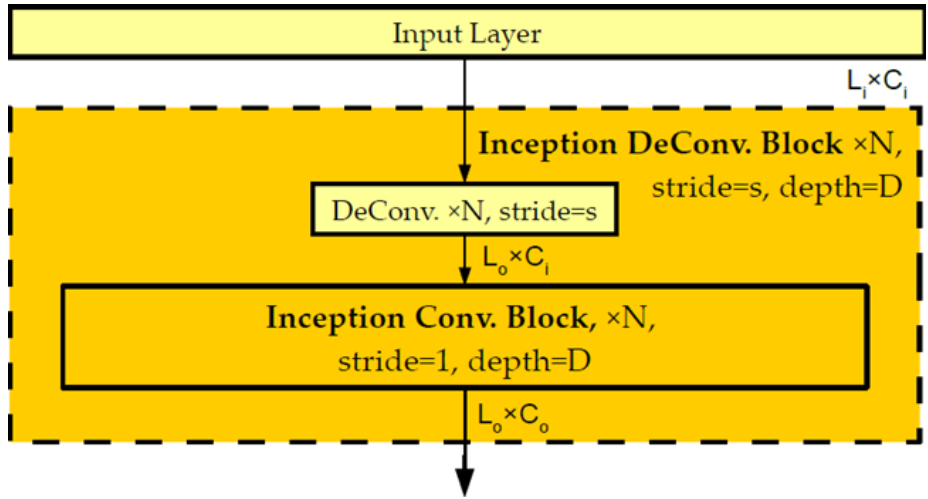
{Basic units of inception network. a) convolutional block; b) deconvolutional block.}
\plot{figure_9}{width=0.5\textwidth}
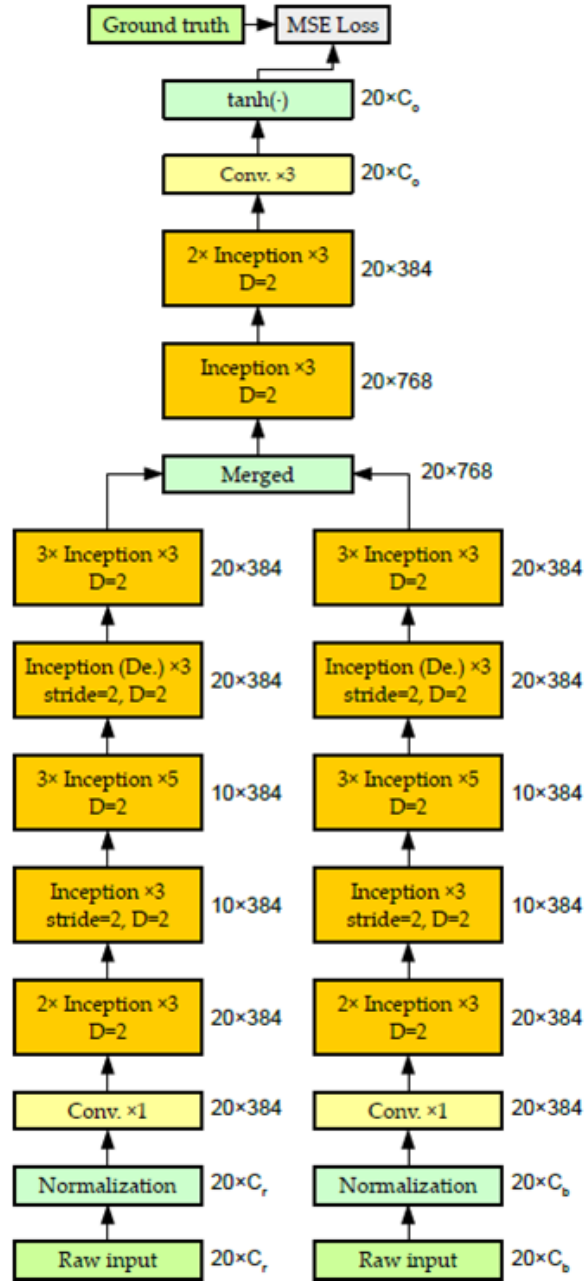
{Structure of the D-depth inception network.}

For a proof of concept, a synthetic example is tested to demonstrate the power of this data-driven method for the LF seismic data prediction. The true velocity model to be inverted for is shown in Figure~\ref{fig:figure_10a}, where the strong reflector imposes a main challenge if the FWI process is initiated by a set of relatively high frequency measurement data and a simple starting velocity model shown in Figure~\ref{fig:figure_10b}. Another velocity model with the same dimension as the true velocity model (Figure~\ref{fig:figure_10c}) is arbitrarily selected to input into a forward modeling simulator to generate a set of synthetic training data. The training dataset contains the HF components ranging from 10 $Hz$ to 18 $Hz$ in 0.5 $Hz$ increments, and an LF component (3 $Hz$), resulting in 18 discrete frequencies in total. After that, the Beat Tone data with $∆\Delta f = 3 \ Hz$ are derived from the HF component pairs in the training dataset (i.e., 10 $Hz$ and 13 $Hz$, 10.5 $Hz$ and 13.5 $Hz$, … etc.). Eventually, both the amplitude-normalized HF training data and the Beat Tone training data are input into the Dual Data Feed structured network depicted in Figure~\ref{fig:figure_7} and Figure~\ref{fig:figure_9} for the supervised learning where the 3 $Hz$ training data serve as the ground truth. The fully trained DNN is able to predict the 3 $Hz$ data with sufficient accuracy from an FWI perspective. Similarly, the 5 $Hz$ data are predicted by repeating the network training and the network testing procedure. To evaluate the accuracy of the LF data predicted by the DNN, the predicted 3 $Hz$ and 5 $Hz$ data are fed into an FWI engine sequentially to perform the velocity model inversion, followed by a series of HF FWI on the 10 $Hz$ to 30 $Hz$ measurement data. The final FWI-reconstructed velocity model in Figure~\ref{fig:figure_11a} is compared against the reference solution shown in Figure~\ref{fig:figure_11b}. The reference solution is produced by the same FWI engine inverting the measurement data covering the full bandwidth from 3 $Hz$ to 30 $Hz$ to avoid the cycle-skipping issue. According to the numerical experiment results demonstrated in Figure~\ref{fig:figure_11a} and Figure~\ref{fig:figure_11b}, the velocity model reconstructed by inverting the predicted LF data is nearly identical to that produced by the true LF data. On the other hand, without LF components, FWI completely fails to resolve the strong reflector in the subsurface although the near surface topography features are successfully recovered, as observed in Figure~\ref{fig:figure_11c}.
%
%
\multiplot{1}{figure_10a,figure_10b,figure_10c}{width=0.8\textwidth}
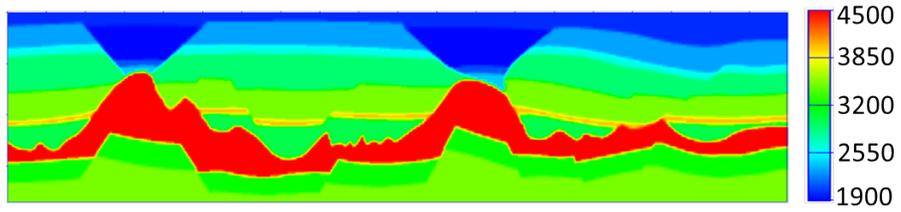
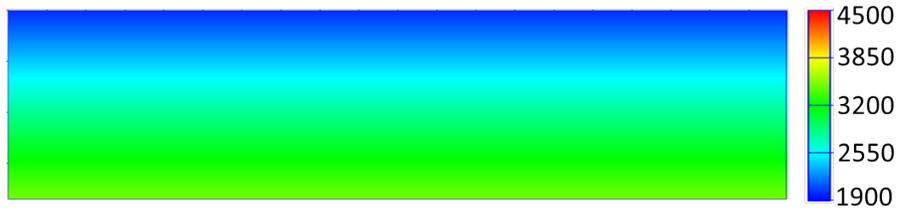
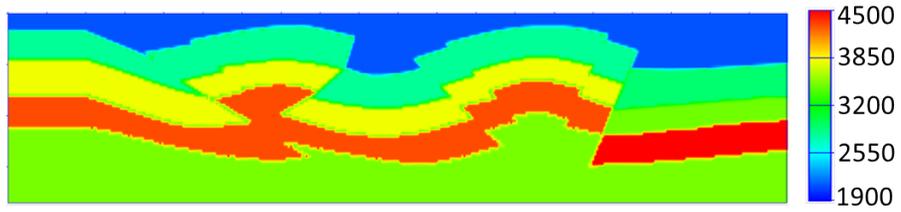
{a) true velocity model; b) initial velocity model for FWI; c) training velocity model.}
%
%
\multiplot{1}{figure_11a,figure_11b,figure_11c}{width=0.8\textwidth}
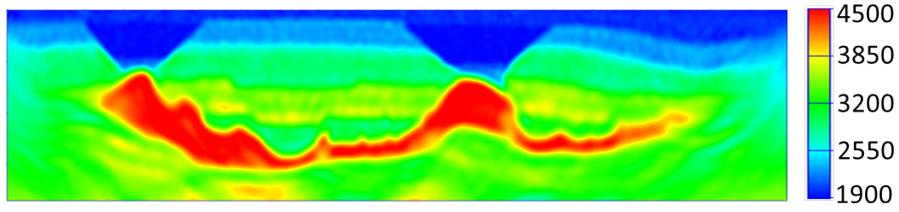
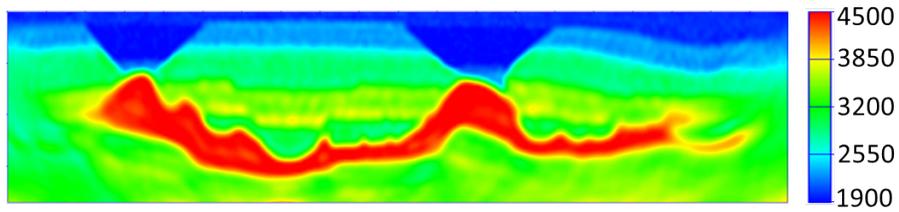
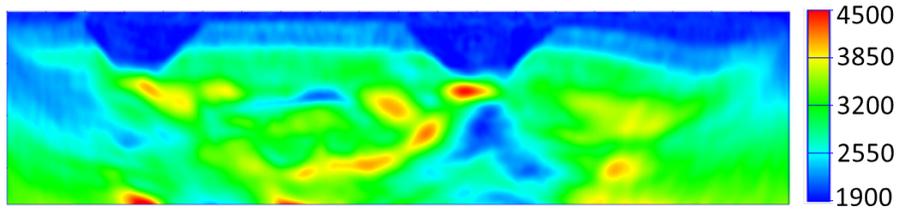
{a) velocity model produced by FWI engine inverting 3 $Hz$ and 5 $Hz$ data predicted by the DNN, followed by FWI inversion of measurement data from 10 $Hz$ to 30 $Hz$. The 3 $Hz$ and 5 $Hz$ data are predicted by learning 10 $Hz$ to 18 $Hz$ measurement data; b) reference solution produced by FWI sequential inversion of  3 $Hz$ to 30 $Hz$ measurement data; c) velocity model produced by FWI sequentially inverting 10 $Hz$ to 30 $Hz$ data.}

\section{Training Velocity Model Selection and Sampling Bias}

The numerical experiment result presented in Figure~\ref{fig:figure_11a} demonstrates the capability of this pure data-driven method to accurately predict the absent LF phase information from the acquired HF data. The predicted LF data are subsequently exploited by a standard FWI engine to successfully reconstruct the strong reflector in the subsurface velocity model, a well-known challenging task. However, one may argue that there is no solid conclusion should be drawn from this numerical experiment regarding the robustness and adaptiveness of this approach, because the training velocity model (Figure~\ref{fig:figure_10a}) and the true velocity model (Figure~\ref{fig:figure_10c}) show similar patterns. Intuitively, an arbitrarily selected training velocity model is non-representative to adequately quantify the nonlinear relationship between the LF data and the HF data, and this arbitrary training data sampling strategy may severely bias the learning process and degrade the prediction accuracy.
%
%
\multiplot{1}{figure_12a,figure_12b,figure_12c}{width=0.8\textwidth}
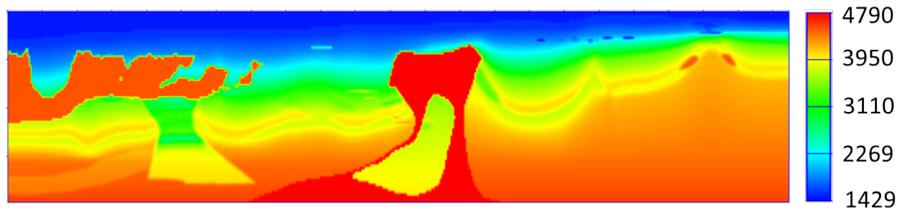
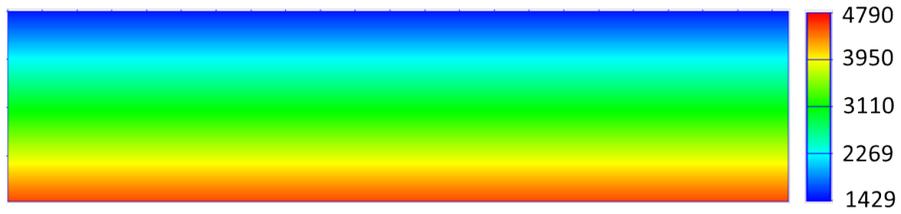
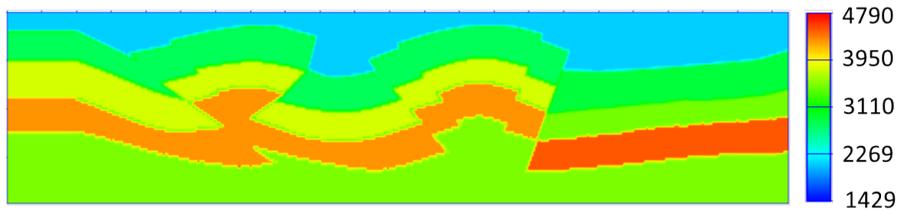
{a) true subsurface velocity model to reconstruct by FWI; b) starting velocity model for FWI; c) training velocity model arbitrarily selected.}

Prompted by this concern, we conducted another numerical experiment on more complex geological structures to investigate the impact of the training velocity model on the DNN performance. The true velocity model for this experiment is shown in Figure~\ref{fig:figure_12a} and the simple 1-D starting velocity model for FWI is displayed in Figure~\ref{fig:figure_12b}. Again, we assume that only HF seismic data beyond 10 $Hz$ are acquired and the primary goal is to predict the phase of LF components (3 $Hz$, 5 $Hz$, and 7 $Hz$). The prediction accuracy is to be evaluated quantitatively in both the data domain (comparing the predicted LF data against the true LF data produced by the forward simulation engine) and the model domain (comparing the velocity model reconstructed by inverting the predicted LF data against the reference FWI solution). A similar training velocity model is selected as shown in Figure~\ref{fig:figure_12c} to initiate the network training. In this experiment, as visually observed in Figure~\ref{fig:figure_12c}, the training velocity model is completely uncorrelated with the true velocity model. A full-bandwidth training dataset with the frequency contents ranging from 3 $Hz$ to 30 $Hz$ is produced by running a forward modeling simulator on the training velocity model. The training curve is plotted in Figure~\ref{fig:figure_13a}, where the loss function reduces to 0.038 after 80 epochs. The absent LF components in the testing dataset are then predicted by the trained network. The evolution of the cross-correlation between the predicted LF data and the true data is plotted in Figure~\ref{fig:figure_13b} as a quantitative measure of the network prediction accuracy. The relatively large error in the predicted LF data, indicated by the cross-correlation value 0.60 of the fully trained network, is presumably induced by the non-representative nature of the training velocity model. This large prediction error is expected to reflect a significant negative impact on the FWI velocity reconstruction. To justify this surmise, the predicted 3 $Hz$, 5 $Hz$, and 7 $Hz$ components are injected into an FWI engine to resolve the large scale structures in the subsurface space. The obtained velocity model is further refined by sequentially inverting the measured HF data ranging from 10 $Hz$ to 30 $Hz$ using the same FWI engine, rendering to the final result shown in Figure~\ref{fig:figure_14}. The predicted LF data successfully help the FWI engine precisely delineate the water-bottom and roughly resolve the top of salt, but fails to properly recover most of the subsurface geophysical features. Strong cycle-skipping-induced artifacts contaminate the velocity model, especially in the sub-salt region at the center of the inversion domain. This numerical experiment confirms our concerns of the non-representative training data and the inherent sampling bias nature of this single-training-model approach, implying that accurate prediction of LF data is unlikely, if not impossible, unless the geological settings are unrealistically simple. Another disadvantage of this DNN method worth mentioning is that there is no effective way to validate the prediction and evaluate the accuracy; thus, applying this approach to real field data projects tends to be a risky practice.
%
%
\multiplot{1}{figure_13a,figure_13b}{width=0.8\textwidth}
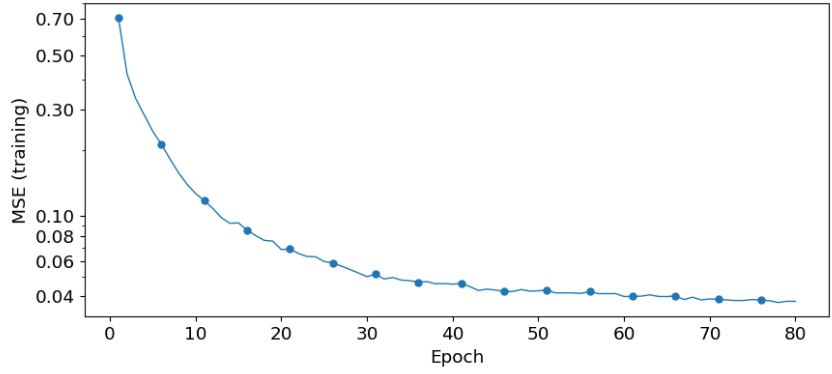
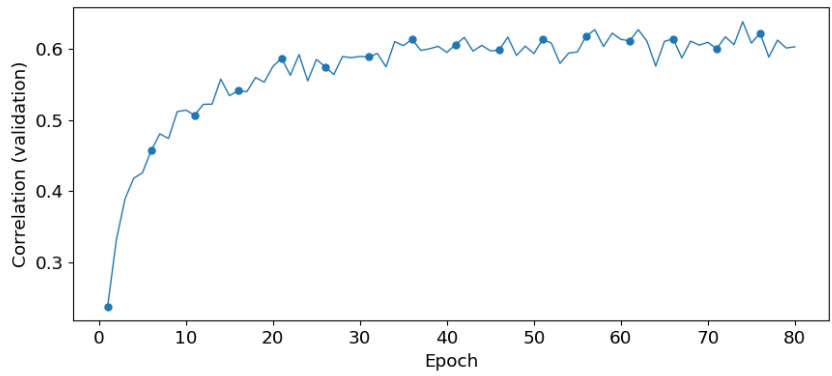
{The performance of first Progressive Transfer Learning iteration. a) learning curve using the training model in Figure~\ref{fig:figure_12c}; b) correlation between the predicted LF data and the true LF data as a measure of the prediction accuracy.}
\plot{figure_14}{width=0.8\textwidth}
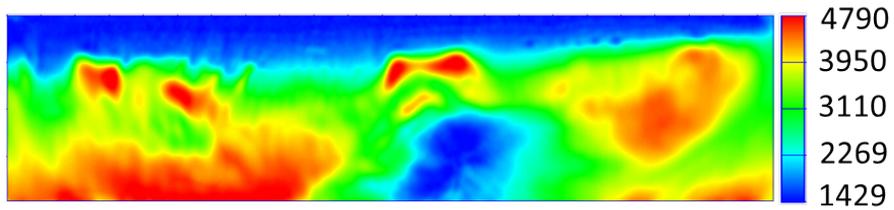
{Reconstructed velocity model by FWI inverting predicted 3 $Hz$, 5 $Hz$, and 7 $Hz$ data, followed by high frequency FWI performed sequentially from 10 $Hz$ to 30 $Hz$.}
\section{Progressive Transfer Learning Method}

Because the root cause of the LF prediction failure in the previous numerical experiment is identified as the non-representative training velocity model, an immediate but cumbersome solution to this problem is a random velocity model generator to represent as many different geological environments as possible. Theoretically, a random velocity model generator is able to generalize the DNN to be adaptive to all types of different scenarios. Unfortunately, there are at least two major shortcomings of this strategy. First, it is unlikely to be successful or manageable to design and train a universal gigantic DNN to capture the global geological and geophysical features by exhaustively learning numerous randomly generated training velocity models. Second, this approach often suffers from underfitting, preventing the network from absorbing the useful information carried by the great amount of training data. Therefore, a more practical and effective approach is desired.

An important phenomenon observed in the previous numerical experiment that could easily be overlooked is that the DNN trained by the true velocity model gives nearly perfect LF prediction. More specifically, if we assign 80\% of the true data as the training data and the remaining 20\% as the testing data, then the trained network is able to predict the 20\% LF testing data with extremely high accuracy. At the first glance, this observation is by no means surprising or useful because the true velocity model is always the unknowns we intend to seek. However, this phenomenon offers us a valuable guideline for the training velocity model design: under the condition that only one training velocity model is allowed, the closer the training model is to the true model, the higher the network prediction accuracy. Consequently, a straightforward strategy inspired by this guideline is to incorporate the subsurface \emph{a priori} information into the training velocity model. We are again facing a dilemma here: on the one hand, accurate subsurface information is not available until a proper FWI process is performed successfully, while on the other hand, an FWI is unlikely to be successful unless the LF data are predicted accurately by feeding the subsurface information into the DNN training process. In light of these thoughts, we propose the so-called Progressive Transfer Learning method to get through this dilemma and eventually enhance the LF prediction accuracy without being overwhelmed by tremendous amount of training velocity models and training data.
\plot{figure_15}{width=0.6\textwidth}
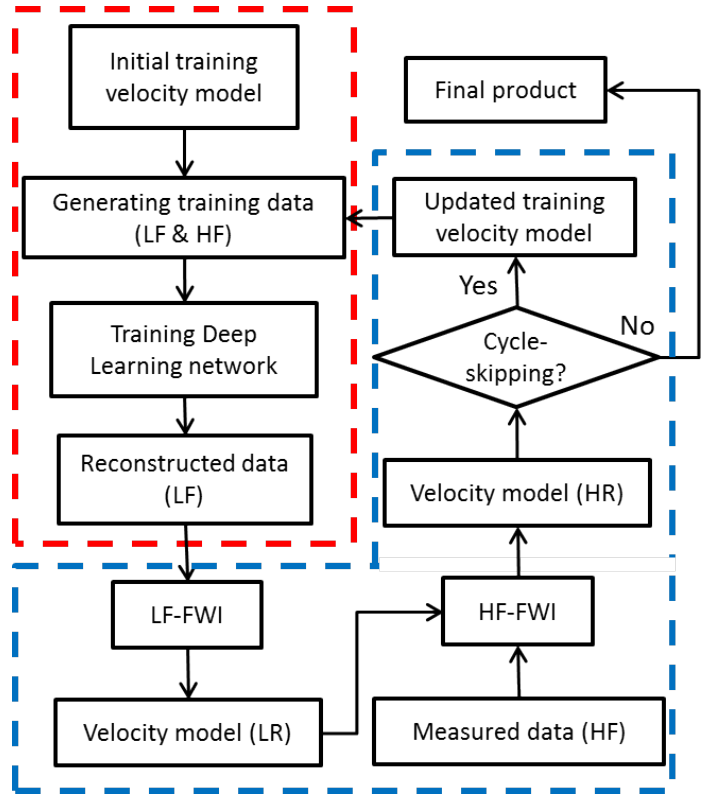
{Workflow of Progressive Transfer Learning for low frequency data reconstruction. LF --- low frequency; HF --- high frequency; LR --- low resolution; HR --- high resolution.}

Rather than throwing all the available training datasets into a gigantic network to ambitiously achieve the adaptiveness to many different varieties of geological and geophysical scenarios, the Progressive Transfer Learning method converts the parallel training to an iterative sequential training process. This method always trains the network using a single training velocity model. Unlike other standard DL approaches, the training dataset in the Progressive Transfer Learning workflow is not fixed but evolves and continuously improves by gradually absorbing more and more reliable subsurface information provided by the physics-based module as the learning process proceeds. With this strategy, the DL network is seamlessly integrated with the physics-based inversion to alternatingly boost each other within every learning cycle.

The workflow of the Progressive Transfer Learning is shown in Figure~\ref{fig:figure_15}, which consists of two main modules: 1) DL module for LF prediction (red block in Figure~\ref{fig:figure_15}); 2) physics-based module (FWI) for training data evolving (blue block in Figure~\ref{fig:figure_15}). The entire learning process is again initiated by an arbitrarily selected training velocity model. After the initial round of network training, the predicted LF data are input into the FWI engine to obtain a low resolution velocity model in the hope of retrieving most of the subsurface low wavenumber structural information. Unfortunately, due to the arbitrarily selected training velocity model, the accuracy of the first round LF data prediction by the network is often insufficient for a successful FWI, as noted in the previous section. Figure~\ref{fig:figure_16a} to Figure~\ref{fig:figure_16c} display three shot gathers (deployed at left, central, and right parts of the domain, respectively) of the predicted 3 $Hz$ components, along with the ground truth. As expected, the accuracy of the LF data prediction varies from shot to shot, and from receiver to receiver. The predicted LF components of the 30th shot (located at the central part of the domain) are in good agreement with the ground truth while the prediction of the 2nd shot (left part of the domain) and 53th shot (right part of the domain) deviate from the ground truth. This observation is counterintuitive because the right part of the true velocity model is relatively less complex because of the lack of salt dome. The FWI inversion of the predicted 3 $Hz$, 5 $Hz$, and 7 $Hz$ data is able to delineate the water-bottom (Figure~\ref{fig:figure_17b}) but fails to resolve the large salt dome in the central region of the domain, which can be clearly identified in FWI result obtained inverting the true LF data (3 $Hz$, 5 $Hz$, 7 $Hz$)(Figure~\ref{fig:figure_17a}).
%
%
\multiplot{1}{figure_16a,figure_16b,figure_16c}{width=0.8\textwidth}
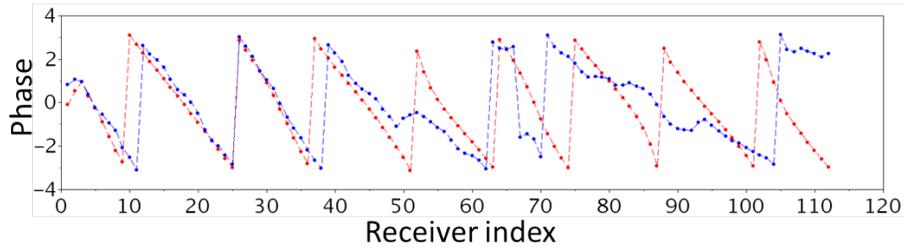
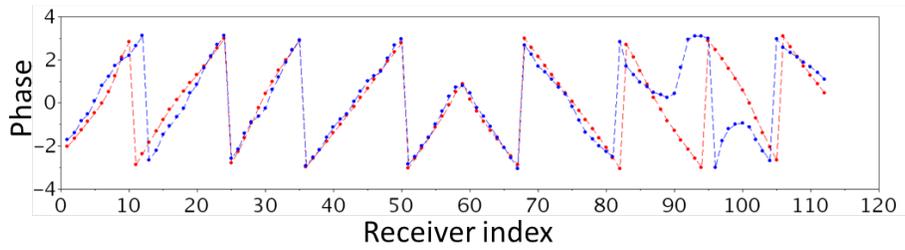
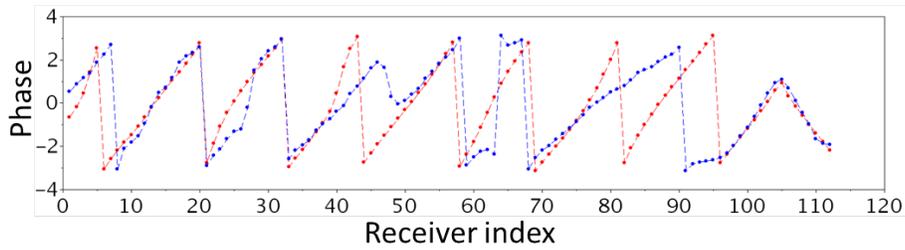
{Comparison between the 1st round transfer learning prediction of 3 $Hz$ data (blue) and the true 3 $Hz$ data (red) for the BP2004 model. a) 2nd shot (deployed at left part of the domain); b) 30th shot (deployed at central part of the domain); c) 53rd shot (deployed at right part of the domain).}
%
%
\multiplot{1}{figure_17a,figure_17b}{width=0.8\textwidth}
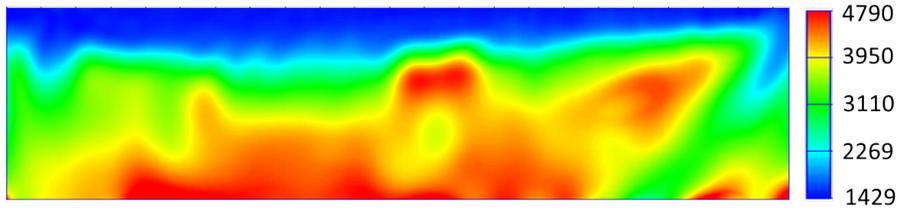
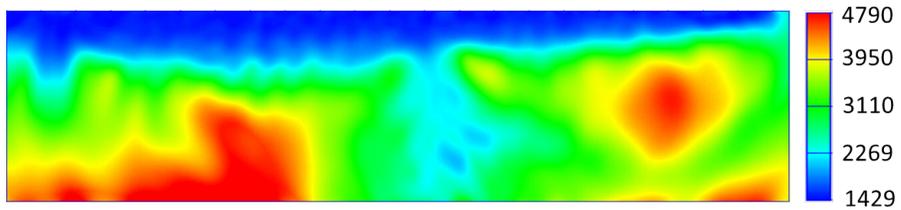
{Comparison between FWI result on LF data. 1) FWI result of true LF data inversion (3 $Hz$, 5 $Hz$, and 7 $Hz$); 2) FWI result of inverting the LF data (3 $Hz$, 5 $Hz$, and 7 $Hz$) predicted by training on an arbitrarily selected velocity model shown in Figure~\ref{fig:figure_12c}.}
%
%
\multiplot{1}{figure_18a,figure_18b,figure_18c}{width=0.8\textwidth}
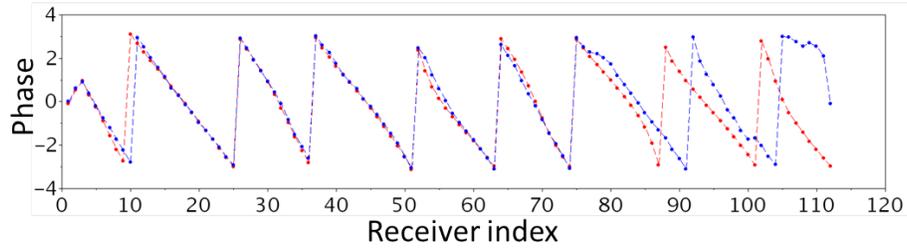
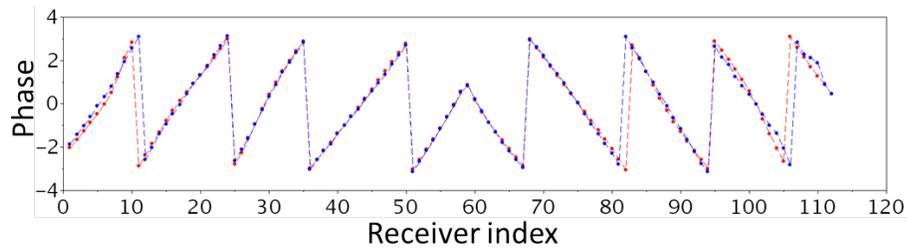
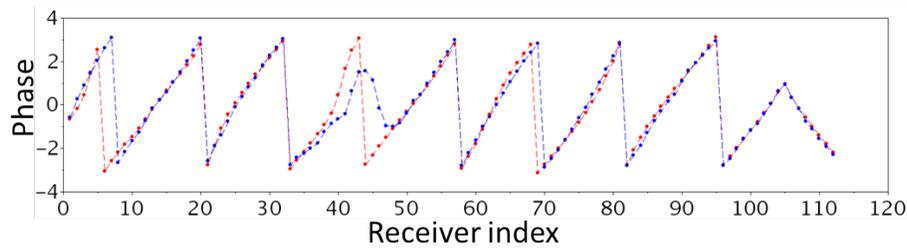
{Comparison between the 2nd round transfer learning prediction of 3 $Hz$ data (blue) and the true 3 $Hz$ data (red) for the BP2004 model. a) 2nd shot (deployed at left part of the domain); b) 30th shot (deployed at central part of the domain); c) 53rd shot (deployed at right part of the domain).}

In spite of the large discrepancy between the prediction and the true LF data, within the framework of Progressive Transfer Learning, the FWI process is continued on the measured HF data up to 30 $Hz$ to complete the first iteration of learning. In other words, the first iteration of the Progressive Transfer Learning is exactly the same approach discussed in the previous section. Therefore, at this early stage of Progressive Transfer Learning, cycle-skipping is expected in the intermediate result shown in Figure~\ref{fig:figure_14}.

Within the scope of Progressive Transfer Learning, the velocity model obtained after the first round LF prediction is sent back into the DL module, acting as an updated training velocity. This strategy is based on two fundamental hypotheses that we do not aim to test rigorously in this work: 1) the machine-learned relationship between the HF and LF components on any training velocity model is able to recover a portion of the subsurface low wavenumber information, while the amount and the reliability of the recovered information is highly dependent on the representative level of the training velocity model; 2) the subsequent FWI operation on the measured HF data is able to amplify and further correct these low wavenumber information because HF data implicitly carry a great amount of such information. Under these two hypotheses, although the velocity model reconstructed after the first round of learning is contaminated by strong artifacts (Figure~\ref{fig:figure_17b}), it contains richer and more accurate low wavenumber components, thus being a more representative and favorable training velocity model than the original one. A new training dataset is then produced by performing a forward simulation on this updated training velocity model to re-train the network, entering the second iteration of Progressive Transfer Learning. The second round LF prediction shows substantial improvement over the previous iteration as plotted in Figure~\ref{fig:figure_18a} to Figure~\ref{fig:figure_18c}, establishing a solid base for the physics-based module to retrieve the subsurface low wavenumber information more reliably. In the every subsequent learning iteration, the DL module always feeds a set of enhanced LF prediction data into the physics-based module. On the other hand, the physics-based module continuously offers a more representative training velocity model and the corresponding enhanced training dataset to the DL module. Thus, the DL module and the physics-based module complement each other alternatingly in an iterative manner, progressively propelling the velocity model inversion process out of the local minima. In this numerical experiment, after three Progressive Transfer Learning cycles, the predicted LF data are free of cycle-skipping and nearly converge to the true LF data, as demonstrated in Figure~\ref{fig:figure_19a} to Figure~\ref{fig:figure_19c}. Similar to the first learning cycle, after 80 epochs of training, the loss function reduces to a small number 0.018, as plotted in Figure~\ref{fig:figure_20a}. While the improvement in the convergence rate of the learning process is subtle, it is important to note that the main benefit brought by the multiple transfer learning cycles is reflected in the LF prediction accuracy, which is quantitatively measured by the cross-correlation validation result plotted in Figure~\ref{fig:figure_20b}. The final cross-correlation value between the predicted LF data and the true LF data is boosted from 0.60 to 0.93 through the three Progressive Transfer Learning cycles, which confirms our forecast that a more representative training velocity model dramatically enhances the LF prediction accuracy. The final FWI result using the predicted LF data (3 $Hz$, 5 $Hz$, and 7 $Hz$) and the true measured HF data is shown in Figure~\ref{fig:figure_21a}. For an informative comparative study, the reference solution obtained by inverting the true full-bandwidth data (3 $Hz$ – 30 $Hz$) and the one produced by inverting the HF data (10 $Hz$ – 30 $Hz$) only are displayed in Figure~\ref{fig:figure_21b} and Figure~\ref{fig:figure_21c}, respectively. This numerical experiment demonstrates that, given a 1-D starting velocity model and an arbitrarily selected simple training velocity model, the Progressive Transfer Learning approach combined with a conventional FWI engine precisely resolves the shallow anomalies and successfully reconstructs the complex salt structures, as one would expect from the excellent agreement between the predicted LF data and the true LF data. On the other hand, the direct inversion of the HF data starting from 10 $Hz$ ends up with severe cycle-skipping-induced artifacts and the completely missing salt structure. The less favorable velocity reconstruction quality at the edges of the domain is probably due to the imbalanced illumination incurred by the offset limitation.
%
%
\multiplot{1}{figure_19a,figure_19b,figure_19c}{width=0.8\textwidth}
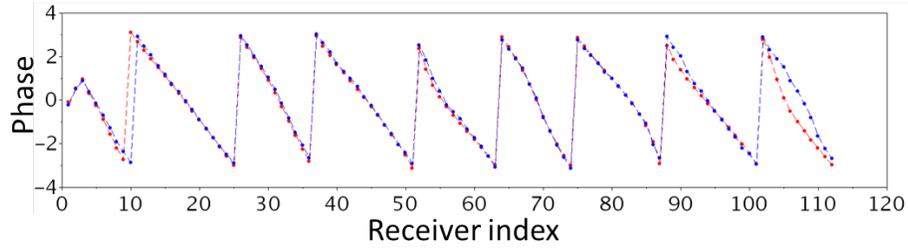
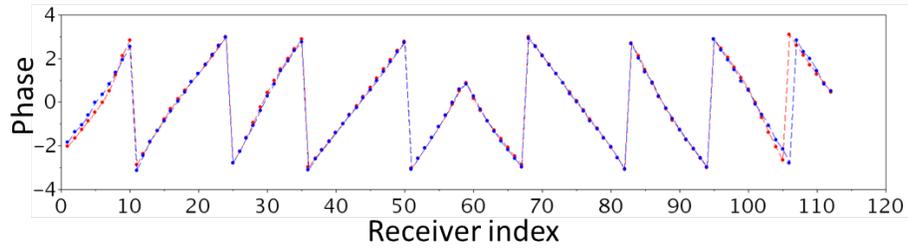
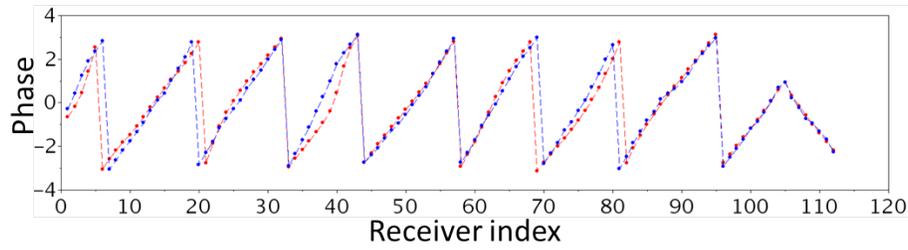
{Comparison between the 3rd round transfer learning prediction of 3 $Hz$ data (blue) and the true 3 $Hz$ data (red) for the BP2004 model. a) 2nd shot (deployed at left part of the domain); b) 30th shot (deployed at central part of the domain); c) 53rd shot (deployed at right part of the domain).}
%
%
\multiplot{1}{figure_20a,figure_20b}{width=0.8\textwidth}
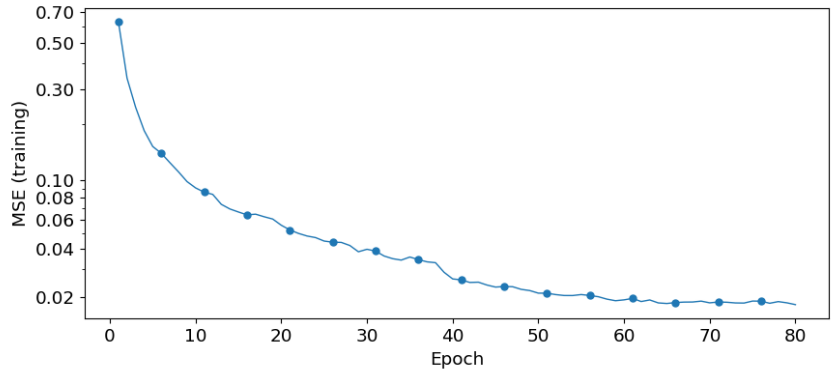
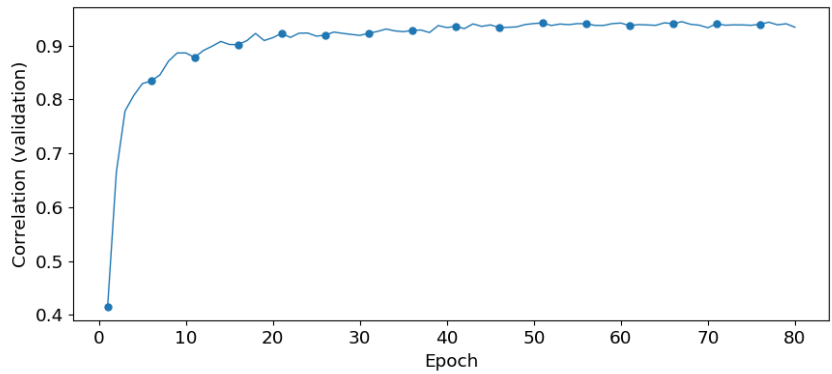
{a) learning curve of the 3rd iteration of Progressive Transfer Learning; b) cross-correlation between the predicted LF data after the 3rd iteration of Progressive Transfer Learning and the true LF data as a measure of prediction accuracy.}
%
%
\multiplot{1}{figure_21a,figure_21b,figure_21c}{width=0.8\textwidth}
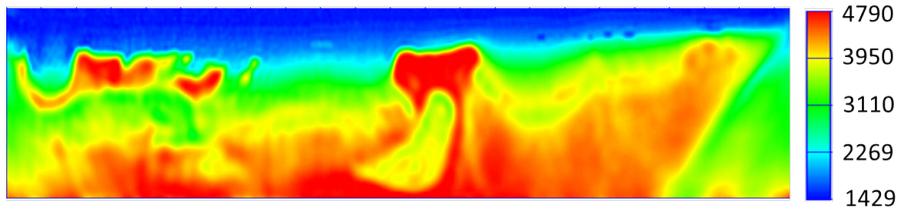
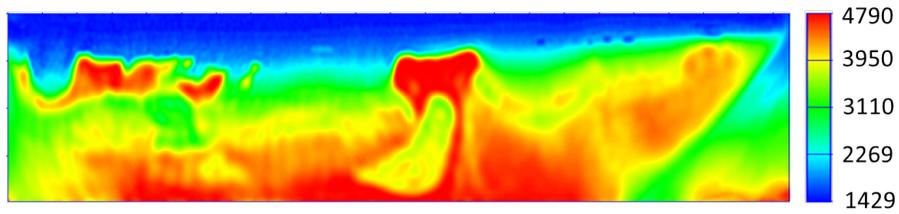
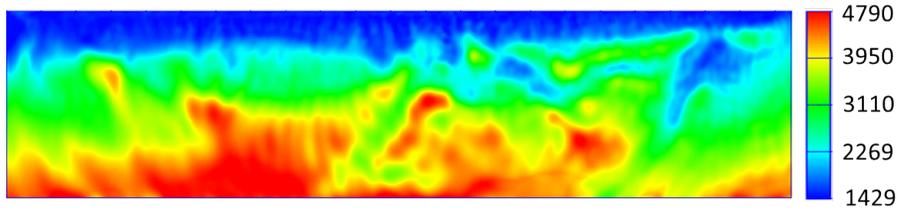
{a) FWI result produced by inverting the 3rd round transfer learning predicted 3 $Hz$, 5 $Hz$, 7 $Hz$ data, followed by measured HF data (10 $Hz$ to 30 $Hz$) inversion; b) FWI result produced by sequentially inverting full-bandwidth measured data ( 3 $Hz$ to 30 $Hz$); c) FWI result produced by direct inverting measured high frequency data sequentially from 10 $Hz$ to 30 $Hz$.}

\section{Conclusion}

In this research work, we developed a novel deep learning based method, the so-called Progressive Transfer Learning, to reconstruct the absent low frequency components in acquired seismic datasets by learning the implicit nonlinear relationship between different frequency bands. This data-driven approach does not require any \emph{a priori} information of the subsurface geological structures and geophysical properties. Instead, the subsurface information is gradually retrieved from the data and pumped into the DNN within the learning process. This Progressive Transfer Learning method contains two main modules: the deep learning module and the physics-based module. After the initial network training on an arbitrarily selected training velocity model, an inaccurate initial low frequency data prediction is performed. In the every subsequent Progressive Transfer Learning iteration, the physics-based module provides an improved training velocity model with richer subsurface information to the deep learning module. On the other hand, the deep learning module updates the low frequency prediction with increased accuracy, and thus, in turn, enables the physics-based module to retrieve more and more reliable subsurface information. With this strategy, the deep learning module and the physics-based module are integrated seamlessly, interacting and complementing with each other to progressively push the FWI process off the local minima. The Progressive Transfer Learning process can be quantitatively monitored because the predicted low frequency data are expected to converge to the training data at the end of a successful transfer learning process, which also serves as a key reliability indicator of the final FWI results. The numerical experiments validate the effectiveness and robustness of the Progressive Transfer Learning method by successfully reconstructing the complex geological structures including multiple salt bodies. The comparative study shows, without the Progressive Transfer Learning Strategy, the same DNN fails to predict the low frequency components with sufficient accuracy in complex geological environment if it is trained by a single velocity model, casting doubts on its practicality on field data projects. Instead of establishing a huge training velocity model library to exhaustively capture the global geological and geophysical characteristics, the Progressive Transfer Learning method extracts local subsurface features through a sequential learning process aided by the physics-based inversion. This unique self-learning feature saves us from being overwhelmed with large amount of training data without sacrificing the prediction accuracy.

\section{Acknowledgement}

This material is based upon work supported by the U.S. Department of Energy, Office of Science, SC-1, under Award Number DE-SC0019665.

\section{Disclaimer}

This report was prepared as an account of work sponsored by an agency of the United States Government.  Neither the United States Government nor any agency thereof, nor any of their employees, makes any warranty, express or implied, or assumes any legal liability or responsibility for the accuracy, completeness, or usefulness of any information, apparatus, product, or process disclosed, or represents that its use would not infringe privately owned rights.  Reference herein to any specific commercial product, process, or service by trade name, trademark, manufacturer, or otherwise does not necessarily constitute or imply its endorsement, recommendation, or favoring by the United States Government or any agency thereof.  The views and opinions of authors expressed herein do not necessarily state or reflect those of the United States Government or any agency thereof.

\newpage

\bibliographystyle{seg}  
\bibliography{DL_LF_FWI}

\end{document}